\documentclass[letters,fleqn,usenatbib]{mnras}

\usepackage{mathptmx}
\usepackage[T1]{fontenc}
\usepackage{ae,aecompl}

\usepackage{graphicx}	% Including figure files
\usepackage{amsmath}	% Advanced maths commands
\usepackage{amssymb}	% Extra maths symbols

\defcitealias{2013ApJ...779...75S}{SM13}

\title[Submm broad RL from NGC 1068]{ALMA observations of the submillimetre hydrogen recombination line from the type 2 active nucleus of NGC 1068}

\author[T. Izumi et al.]{
Takuma Izumi,$^{1}$\thanks{E-mail: takumaizumi@ioa.s.u-tokyo.ac.jp}
Kouichiro Nakanishi,$^{2,3}$
Masatoshi Imanishi,$^{4,2,3}$
and Kotaro Kohno$^{1,5}$
\\
$^{1}$Institute of Astronomy, School of Science, The University of Tokyo, 2-21-1 Osawa, Mitaka, Tokyo 181-0015, Japan\\
$^{2}$National Astronomical Observatory of Japan, 2-21-1 Osawa, Mitaka, Tokyo 181-8588, Japan\\
$^{3}$SOKENDAI (The Graduate University for Advanced Studies), 2-21-1 Osawa, Mitaka, Tokyo 181-8588, Japan\\
$^{4}$Subaru Telescope, NAOJ, 650 North A'ohoku Place, Hilo, HI 96720, USA\\
$^{5}$Research Center for the Early Universe, The University of Tokyo, 7-3-1 Hongo, Bunkyo, Tokyo 113-0033, Japan
}

\date{Accepted 2016 February 08. Received 2016 January 30; in original form 2016 January 13}

\pubyear{2015}

\begin{document}
\label{firstpage}
\pagerange{\pageref{firstpage}--\pageref{lastpage}}
\maketitle

% Abstract of the paper
\begin{abstract}
Hydrogen recombination lines at the submillimetre band (submm-RLs) can serve as probes of ionized gas without dust extinction. 
One therefore expects to probe the broad line region (BLR) of an obscured 
(type 2) active galactic nucleus (AGN) with those lines. 
However, admitting the large uncertainty in the continuum level, 
here we report on the non-detection of both broad and narrow H26$\alpha$ emission line 
(rest frequency = 353.62 GHz) towards the prototypical type 2 AGN of NGC 1068 
with the Atacama Large Millimeter/submillimeter Array (ALMA). 
We also investigate the nature of BLR clouds that can potentially emit submm-RLs with model calculations. 
As a result, we suggest that clouds with an electron density ($N_e$) of $\sim$ 10$^9$ cm$^{-3}$ 
can mainly contribute to broad submm-RLs in terms of the line flux. 
On the other hand, line flux from other density clouds 
would be insignificant considering their too large or too small line optical depths. 
However, even for the case of $N_e \sim 10^9$ cm$^{-3}$ clouds, 
we also suggest that the expected line flux is extremely low, 
which is impractical to detect even with ALMA. 
\end{abstract}

% Select between one and six entries from the list of approved keywords.
% Don't make up new ones.

%%%%%%%%%% SECTION 1: INTRODUCTION %%%%%%%%%%
\section{Introduction}\label{sec1}
Since the proposition of the unification scheme of an active galactic nucleus (AGN; \citealt{1993ARA&A..31..473A}), 
it has been supposed that a type 2 AGN contains a type 1 nucleus 
having a broad line region (BLR) of ionized material, 
which is obscured by a dusty torus at short wavelengths such as optical. 
The detection of the broad \ion{H}{i} recombination line (RL, hereafter) in the polarized light 
from the prototypical type 2 Seyfert NGC 1068 (e.g., \citealt{1985ApJ...297..621A,1999MNRAS.304L...1A}) 
constitute the basis of this classical model. 
On the other hand, direct detection of the BLR components 
at the wavelengths least influenced by the dust attenuation 
will be another way to better understand the unification mechanism. 
Although some works succeeded in detecting relatively broad ($\sim$ 1000--3000 km s$^{-1}$) RLs 
(e.g., \citealt{1994ApJ...422..521G,1997ApJ...484...92V,1999ApJ...522..113V,2002A&A...396..439L}) 
in type 2 Seyfert galaxies at near-infrared (NIR), 
these detections seem to be biased towards less-obscured type 2 
or intermediate type AGNs (e.g., \citealt{2000ApJ...530..733L,2002A&A...396..439L}). 
Here, the ability of penetrating columns of $N_{\rm H}$ $\la$ 10$^{23-24}$ cm$^{-2}$ 
expected for the NIR band (e.g., \citealt{1997MNRAS.288..977A}) 
would be insufficient to probe a sizeable fraction of Compton-thick AGNs ($N_{\rm H}$ $>$ 10$^{24}$ cm$^{-2}$). 
Indeed, previous studies failed to detect NIR broad RLs firmly 
from such Compton-thick objects (e.g., \citealt{1994ApJ...422..521G,2000ApJ...530..733L,2010RAA....10..427C,2015ApJS..217...13M}). 
This problem can be severer in probing extremely obscured nuclei whose energy sources are elusive 
($N_{\rm H}$ $\ga$ 10$^{25-26}$ cm$^{-2}$, e.g., \citealt{2013ApJ...764...42S,2015A&A...584A..42A}). 
However, at least in the nearby universe, hard X-ray observations suggest that Compton-thick AGNs 
are comparable or even larger in number density than less obscured AGNs 
(e.g., \citealt{1999ApJ...522..157R,2011ApJ...728...58B,2015ApJ...809..115L}). 
Therefore, longer wavelengths observations that directly probe those heavily obscured BLRs 
are highly desired to investigate its nature as well as 
to examine the applicability of the unified scheme to such an extreme population 
(e.g., \citealt{2015ApJ...802...89B,2015ApJ...806..127D}). 

From this perspective, we considered that submillimetre \ion{H}{i} RLs (submm-RLs, hereafter) 
would provide a way as they do not suffer from dust extinction. 
As is not the case of centimetre RLs, 
submm-RLs would not be affected by potential maser amplification (\citealt{1990ApJ...365..606G}). 
Moreover, it is easy to achieve quite high-velocity resolution at submillimetre 
that enables us to study kinematics of the nuclear regions in detail. 
One critical issue has been the intrinsically very faint nature of the submm-RLs, 
but this would now be overcome thanks to the advent of The Atacama Large Millimeter/submillimeter Array (ALMA), 
which provides unprecedented sensitivity and resolution. 
Indeed, Scoville \& Murchikova (2013, hereafter \citetalias{2013ApJ...779...75S}) predicted 
that both thermal free$-$free continuum and submm-RL emissions from narrow line regions (NLRs) 
of luminous nearby galaxies are readily detectable with ALMA. 
It is therefore the time to test the detectability of submm-RLs 
of the BLR-origin with the actual ALMA data. 

In this paper, we present our ALMA Cycle 2 observations of the submm-RL, 
H26$\alpha$ (rest frequency $\nu_{\rm rest}$ = 353.62 GHz), 
towards the centre of NGC 1068 ($D$ = 14.4 Mpc; \citealt{1988ngc..book.....T}). 
This galaxy hosts a Compton-thick AGN 
($N_{\rm H}$ $\sim$ 10$^{25}$ cm$^{-2}$; e.g., \citealt{2016MNRAS.456L..94M}), 
thus is one of the best targets to demonstrate the capability 
of submm-RLs to study heavily dust-obscured nuclei. 
Note that several attempts have been done so far towards this AGN to detect broad \ion{H}{i} RLs: 
\citet{2000ApJ...530..733L} observed Br$\beta$ 2.63 $\mu$m and Pf$\alpha$ 7.46 $\mu$m emission 
but could not detect broad lines likely due to severe dust extinction. 
No significant emission of H40$\alpha$ ($\nu_{\rm rest}$ = 99.02 GHz) 
and H53$\alpha$ ($\nu_{\rm rest}$ = 42.95 GHz) lines were detected with single dish measurements as well 
at least partly due to insufficient sensitivity (\citealt{1991MNRAS.248..585P}).

%%%%%%%%%% SECTION 2: DATA %%%%%%%%%%
\section{Data analysis}\label{sec2}
We observed NGC 1068 with ALMA on 2015 June 15 with 37 antennas 
as a Cycle 2 science program (ID = 2013.1.00188.S, PI: M. Imanishi). 
The baseline length spans from 21.4 to 783.5 m, which corresponds to the $uv$ range of 25.0 to 914.2 k$\lambda$ at 350 GHz. 
The nuclear region of NGC 1068 was entirely covered with a single pointing with an 18 arcsec field of view. 
Throughout the work, the location of the AGN is set to ($\alpha_{\rm J2000.0}$, 
$\delta_{\rm J2000.0}$ = 02$^{\rm h}$42$^{\rm m}$40$\fs$710, $-$00$^\circ$00$'$47$\farcs$938), 
which is defined by high-resolution interferometric observations at 6 cm (e.g., \citealt{2004ApJ...613..794G}). 
The receiver was tuned to cover the redshifted HNC($J$ = 4--3) line 
($\nu_{\rm rest}$ = 362.63 GHz) in the upper side band (USB), 
and the redder side of H26$\alpha$ \ion{H}{i} RL ($\nu_{\rm rest}$ = 353.62 GHz) 
relative to the systemic velocity ($V_{\rm sys}$ = 1137 km s$^{-1}$; \citealt{1999ApJS..121..287H}) in the lower side band (LSB). 
Each of four spectral windows (Spw 0, 1 = USB and Spw 2, 3 = LSB) has 
a bandwidth of 1.875 GHz, which yields $\sim$ 7.5 GHz width in total. 
Note that we could not cover the entire velocity range of the hypothesized broad H26$\alpha$ 
because the LSB has only $\sim$ 3180 km s$^{-1}$ width in total for this line. 
In this paper, we focus on this H26$\alpha$ line and the underlying continuum emissions only. 

The velocity spacing of the LSB was originally 0.42 km s$^{-1}$ per channel, 
but 120 channels were binned to improve the signal-to-noise ratio, 
which resulted in the final velocity resolution of d$V$ $\sim$ 50 km s$^{-1}$. 
This is high enough to resolve the hypothesized broad line component, 
which has a full width at zero intensity (FWZI) of $\sim$ 7500 km s$^{-1}$ 
as seen in the polarized spectrum of H$\beta$ (\citealt{1985ApJ...297..621A}). 
It is also enough to moderately resolve the typical width of NLR-lines of NGC 1068 
($\sim$ 500 km s$^{-1}$; \citealt{2011ApJ...739...69M}). 
The bandpass, phase, and flux were calibrated with J0224+0659, J0239+0416, and Ceres, respectively. 
The total on-source time was $\sim$ 35 min. 

The reduction, calibration, and subsequent analysis were conducted with \verb|CASA| 
version 4.3.0 (\citealt{2007ASPC..376..127M}) in standard manners. 
All images were reconstructed by the \verb|CASA| task \verb|CLEAN|. 
Continuum subtraction can be problematic in the LSB because all channels 
would contain the BLR component due to the limited band width. 
Therefore, we show the spectrum of the LSB without continuum subtraction in this work. 
The information on the continuum emission is extracted from the USB instead (Section \ref{sec3}). 
The full width at the half-maximum (FWHM) of the achieved synthesized beam was 
$\theta_{\rm maj}$ $\times$ $\theta_{\rm min}$ = 0.37 arcsec $\times$ 0.31 arcsec, 
which corresponds to 26 pc $\times$ 22 pc at the distance of NGC 1068. 
The 1$\sigma$ sensitivity is 0.42 mJy beam$^{-1}$ 
at d$V$ = 50 km s$^{-1}$ for the LSB (estimated from the areas free from emission), 
whereas 1$\sigma$ = 0.19 mJy beam$^{-1}$ for the USB continuum emission, 
which is surely free from the contamination of the broad H26$\alpha$ line.

%%%%%%% SECTION 3: RESULTS %%%%%%%%%%%%
\section{Results}\label{sec3}
\begin{figure}
	\includegraphics[width=\columnwidth]{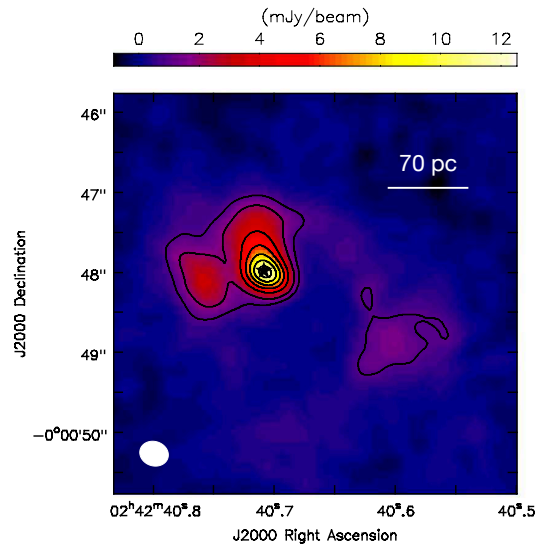}
    \caption{The USB continuum emission of the central region 
    of NGC 1068 centred at $\nu_{\rm rest}$ = 364 GHz. 
    The white ellipse denotes the synthesized beam of 0.37 arcsec $\times$ 0.31 arcsec 
    (1 arcsec = 70 pc) with P.A. = 73$\fdg$64. 
    The contour indicates 5, 10, 20, $\ldots$, and 60$\sigma$, where 1$\sigma$ = 0.19 mJy beam$^{-1}$. 
    The flux density at the AGN position (central star; \citealt{2004ApJ...613..794G}) is 11.72 mJy beam$^{-1}$.
    }
    \label{fig1}
\end{figure}

\begin{figure}
	\includegraphics[width=\columnwidth]{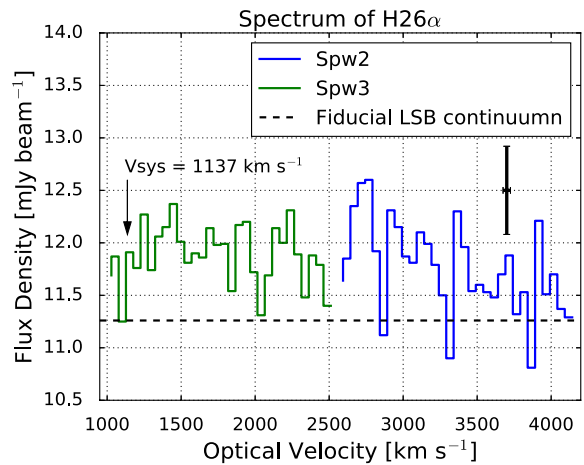}
    \caption{The LSB spectrum ($\nu_{\rm rest}$ = 350.08--353.85 GHz) extracted at the AGN position of NGC 1068. 
    This frequency range must contain the H26$\alpha$ emission from the BLR. 
    The vertical solid line shows the $\pm$1$\sigma$ sensitivity (0.42 mJy per 0.37 arcsec $\times$ 0.31 arcsec beam; 1 arcsec = 70 pc) 
    at the velocity resolution of 50 km s$^{-1}$. 
    The label of the $x$-axis is expressed as the LSR velocity of H26$\alpha$ in the optical convention. 
    The systemic velocity of NGC 1068 is indicated by the arrow, 
    whereas the horizontal dashed line denotes our fiducial estimate of the LSB continuum level (11.26 mJy; see the text for details). 
    }
    \label{fig2}
\end{figure}

Fig. \ref{fig1} shows the spatial distribution of the USB continuum emission 
centred at $\nu_{\rm rest}$ = 364 GHz towards the nuclear region of NGC 1068. 
This distribution is well consistent with that of the 349 GHz continuum emission (\citealt{2014A&A...567A.125G}). 
The emission is clearly detected and peaks at the precise AGN position 
(11.72 mJy beam$^{-1}$), which suggests this is of AGN origin. 
Fig. \ref{fig2} displays the LSB spectrum extracted at the AGN position with a single aperture. 
The plotted range corresponds to almost the redder half of the FWZI 
of the polarized broad H$\beta$ (\citealt{1985ApJ...297..621A}). 

Next, we estimate the continuum level at the LSB using that at the USB. 
So far, three models have been mainly proposed to explain the centimetre 
to submillimetre nuclear spectral energy distribution (SED) of NGC 1068 
(see \citealt{2011ApJ...736...37K} and references therein): ($i$) pure synchrotron model, 
($ii)$ thermal free$-$free model\footnote{The origin of this free$-$free emission 
was modelled (\citealt{2008A&A...485...33H}) 
to have a modest electron density (8 $\times$ 10$^5$ cm$^{-3}$) 
and a very high temperature (1.3 $\times$ 10$^6$ K). 
Hence, the dominant origin would not be the BLR clouds.}, 
and ($iii$) electron-scattered synchrotron model. 
These models consider a significant contribution 
from thermal dust emission as well. 
Among them, previous studies supported the models ($ii$) and ($iii$) equally 
(e.g., \citealt{2011ApJ...736...37K,2014A&A...567A.125G}), but disfavoured the model ($i$). 
However, we find that even these preferred two models overestimate our USB continuum 
level by $\sim$ 2 times, judging from their proposed formulations. 
Indeed, the lack of high-resolution photometry at far-infrared, 
as well as different apertures used in previous works at centimetre to submillimetre, 
altogether make the accurate SED modelling quite difficult. 
A highly time variable nature of AGN-originated emission 
even at millimetre to submillimetre (e.g., \citealt{2015MNRAS.454.4277B}) also hinders such a challenge. 

We then simply give a crude estimate on the LSB continuum level 
by assuming that the USB continuum emission is composed of 
(1) thermal dust emission from both the dusty torus and the extended cold dust 
(50 per cent contribution to the total USB emission), 
(2) thermal free$-$free emission from both the BLR and the NLR (25 per cent), 
and (3) synchrotron emission from the jet (25 per cent). 
Indeed, \citet{2014A&A...567A.125G} suggested that $\sim$ 50 per cent 
continuum at the ALMA Band 7 stems from non-dust components. 
Assuming typical slopes ($\alpha$) of the flux scaling ($flux$ $\propto$ $\nu^\alpha$) 
that are 3.5, $-$0.15, $-$1.0 for the above three components, respectively (e.g., \citealt{2011piim.book.....D}), 
we estimate the LSB continuum level to be 11.26 mJy beam$^{-1}$. 
Note that varying the free$-$free to synchrotron ratio only slightly modify this estimation, 
which does not influence the following discussion. 
Moreover, the free-free continuum levels at the USB and the LSB are almost identical 
irrespective of its contribution to the total continuum flux and its spectral index. 

Back to Fig. \ref{fig2} with keeping this fiducial continuum level in mind, 
one would see that there seems to be some excess in the flux over a wide velocity range, 
which can be the broad H26$\alpha$ emission line. 
However, this excess in each channel is still below 
the 3$\sigma$ limit of our observations (1.26 mJy beam$^{-1}$). 
Moreover, the amount of the excess strongly depends on the actual continuum level, 
which is hard to determine at this moment. 
Considering this situation, we call this marginal excess at the LSB as {\it{non-detection}}. 
Although this view is not conclusive, 
we nevertheless summarize our current estimates in Table \ref{tbl1}. 

Although it is not detected convincingly, the potential existence of the broad submm-RL 
is an appealing topic to investigate. 
Therefore, in the following, we discuss the nature of the hydrogen clouds 
that can actually emit broad H26$\alpha$ line with invoking simple theoretical calculations, 
as well as the detectability of submm-RLs with future high-sensitivity observations. 
These predictions can be tested with future high-resolution, 
high-sensitivity, and quasi-synchronized observations 
over a velocity range enough to cover the entire 
(hypothesized) broad H26$\alpha$ line emission. 

\begin{table*}
	\centering
	\caption{Emission properties at the centre of NGC 1068}
	\label{tbl1}
	\begin{tabular}{lccccc}
		\hline
		Emission & $\nu_{\rm rest}$ & Peak flux & Peak flux & $S$ & $S$ \\
		 & (GHz) & (mJy beam$^{-1}$) & (mK) & (Jy beam$^{-1}$ km s$^{-1}$) & (K km s$^{-1}$) \\
		\hline
		H26$\alpha$ (red side) & 353.62 & $<$1.26 & $<$107 & $<$4.02 & $<$342 \\
		Continuum (USB) & 363.98 & 11.26$\pm$0.19 & 956$\pm$15 & $-$ & $-$ \\
		\hline
	\end{tabular}
	\\{{\bf{Note.}} For H26$\alpha$, we list 3$\sigma$ upper limit on the peak values 
	that are appropriate for both the broad and the narrow components at the velocity resolution of 50 km s$^{-1}$. 
	The integrated intensity was calculated by assuming 
	the Gaussian profile with the FWHM of 3000 km s$^{-1}$ (\citealt{1999MNRAS.304L...1A}), 
	having the peak flux of the above 3$\sigma$ value.}
\end{table*}

%%%%%%%%%% SECTION 4: DISCUSSION %%%%%%%%%%
\section{Discussion}\label{sec4}
Since the electron volume density ($N_e$) and line optical depths are extremely large in BLRs 
compared to those in typical \ion{H}{ii} regions, 
the so-called Case-B approximation would not be appropriate in a strict sense. 
However, previous studies support the fairy good resemblance between 
the observed line flux ratios and the theoretical values under the Case-B condition 
(see \citealt{1990agn..conf...57N} and references therein). 
Therefore, we adopted the model-calculated parameters of \ion{H}{i} RLs 
under the Case-B condition by \citet{1995MNRAS.272...41S} 
as our best estimate at first. 
Their values were used in \citetalias{2013ApJ...779...75S} 
as well for the cases of $N_e$ $\leq$ 10$^8$ cm$^{-3}$. 
Then, we expand the theoretical basis by \citetalias{2013ApJ...779...75S} towards higher $N_e$ in this work. 
We investigate several cases of a gas cloud with uniform density 
and temperature as BLRs likely to have clouds with various physical conditions. 
Note that we use $N$ to express the volume density to avoid 
confusion with the principal quantum number $n$. 

\begin{figure}
	\includegraphics[width=\columnwidth]{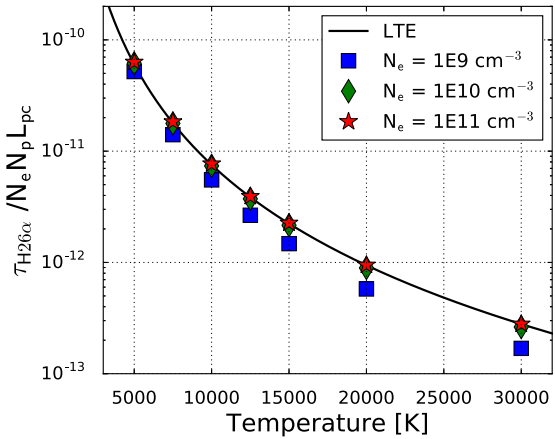}
    \caption{The optical depth of H26$\alpha$ per unit EM, $\Omega_{\rm H26\alpha}$ in Equation (\ref{eq1}), 
    for electron temperature ($T_e$) = 5000$-$30000 K. 
    Three cases of electron density, $N_e$ = 10$^{9}$, 
    10$^{10}$, and 10$^{11}$ cm$^{-3}$ (typical values expected for BLRs; \citealt{1997iagn.book.....P}), 
    as well as the LTE solution are plotted. 
    No significant difference is found among these values. 
    The line centre optical depth can be achieved by multiplying this value 
    by an EM = $\int N_e N_p$d$L$ = $N_e$$N_{\rm p}$$L_{\rm pc}$, 
    where $L_{\rm pc}$ is the line-of-sight path length in the unit of parsec. 
    }
    \label{fig3}
\end{figure}

At the high densities suggested in BLRs ($N_e$ $\ga$ 10$^8$ cm$^{-3}$), one can expect level populations are 
in the local thermal equilibrium (LTE) especially for large $n$. 
Indeed, the departure coefficient from the LTE ($b_n$) calculated by \citet{1995MNRAS.272...41S} 
clearly shows $\sim$ 1 at $N_e$ $\ga$ 10$^8$ cm$^{-3}$ for $n$ $\ga$ 20 
(a range containing submm-RLs), i.e., no departure from the LTE. 
Fig. \ref{fig3} shows the optical depth parameter $\Omega_{n,n'}$ of H26$\alpha$ 
calculated by \citet{1995MNRAS.272...41S} as a function of $N_e$ and electron temperature ($T_e$), 
which is related to the line centre optical depth as 
\begin{equation}
    \tau_{n,n'} = N_e N_p \Omega_{n,n'} L,
	\label{eq1}
\end{equation}
where $L$ is the line-of-sight path length (\citealt{1987MNRAS.224..801H}). 
Thus, the $\Omega_{n,n'}$ gives the optical depth per unit emission measure (EM), 
which is defined as 
\begin{equation}
    \rm{EM} = \int N_e N_p dL.
	\label{eq2}
\end{equation}
$N_p$ is the proton volume density. 
The thermal Doppler width ($\sim$ 21.5 km s$^{-1}$ at $T_e$ = 10$^4$ K for \ion{H}{i}) is assumed in Fig. \ref{fig3} 
as we primarily treat a single cloud existing in BLRs. 
Although this width shall be regarded as just the first-order approximation (e.g., \citealt{1997iagn.book.....P,1999A&A...351...31D}), 
we suppose that a line width up to $\sim$ a few 100 km s$^{-1}$ would not alter our conclusions significantly. 
Note that $\tau_{n,n'}$ is inversely proportional to the line width. 

As is expected from the quasi-LTE condition, we can see from Fig. \ref{fig3} 
that the line optical depth is approaching to the following LTE solution under the ionization equilibrium: 
\begin{equation}
   \tau_{\rm line,LTE} = 2.7 \times 10^3 \times \left( \frac{T_e}{\rm K} \right)^{-3} \times \left( \frac{\rm{EM}}{\rm cm^{-6} ~pc} \right) \times \left( \frac{\nu}{\rm GHz} \right)^{-1}.
	\label{eq3}
\end{equation}
Hence, we will discuss under the LTE condition hereafter. 
Substituting typical values supposed for BLRs 
(e.g., \citealt{1992ApJ...387...95F,1997iagn.book.....P,1999A&A...351...31D,2006agna.book.....O}), 
namely, $N_e$ = $N_p$ = 10$^9$ cm$^{-3}$, 
$T_e$ = 10$^4$ K, and $L$ = 20 r$_\odot$ = 4.5 $\times$ 10$^{-7}$ pc 
(the path length of a single cloud; radius = 10 r$_\odot$ in this case), 
we obtain an optical depth of H26$\alpha$ as $\tau_{\rm H26\alpha}$ $\sim$ 3.4, 
i.e., the line emission is (moderately) optically thick in this case. 
This, reflecting the quite high EM in BLRs adopted here, is a striking difference 
from the cases in \ion{H}{ii} regions, 
where the submm-RLs are always optically thin. 

On the other hand, in the thermal plasma with $T_e$ $\sim$ 10$^4$ K, 
the thermally averaged gaunt factor (\citealt{1980afcp.book.....L}) is, 
\begin{equation}
   \bar{g}_{\rm ff} = \frac{\sqrt{3}}{\pi} \left[ \ln \frac{(2k_{\rm B} T_e)^{3/2}}{4.23 \pi {\rm e}^2 \nu m^{1/2}_p} \right] = 0.55 \left[ 17.7 + \ln \left( \frac{T^{3/2}_e}{\nu} \right) \right] .
	\label{eq4}
\end{equation}
With this formula, an optical depth of the free$-$free emission is 
\begin{equation}
   \tau_{\rm ff} = 0.082 \times \left( \frac{T_e}{\rm K} \right)^{-1.35} \times \left( \frac{\nu}{\rm GHz} \right)^{-2.1} \times \left( \frac{{\rm EM}}{\rm{cm^{-6} ~pc}} \right).
	\label{eq5}
\end{equation}
Substituting the above-adopted parameters, 
we obtain $\tau_{\rm ff}$ $\sim$ 0.65 at 350 GHz for the BLR emission, 
i.e., thermal free$-$free emission is moderately optically thin. 
Therefore, we can expect a relatively high peak line flux to continuum ratio, 
\begin{equation}
R_{\rm lc} \equiv \frac{T_e (1-{\rm e}^{-(\tau_{\rm line} + \tau_{\rm ff})})}{T_e (1-{\rm e}^{-\tau_{\rm ff}})} - 1
\label{eq6}
\end{equation}
of $\sim$ 1.0 in this case. 
We used the formalism by, e.g., \citet{1978ApJ...223..378B}, for this estimation. 
A potential contamination due to stimulated emission caused by 
the background non-thermal radiation is neglected for simplicity. 
The high $R_{\rm lc}$ $>$ 0 seems to open a way to detect the broad submm-RL, 
if clouds with $N_e$ $\sim$ 10$^9$ cm$^{-3}$ 
are the dominant contributor to the BLR flux at submillimetre. 
Moreover, as both $\tau_{\rm H26\alpha}$ 
and $\tau_{\rm ff}$ are functions of $T_e$ and EM, 
we can estimate one parameter when the other is known independently. 
However, that situation is disfavoured in an actual observational sense as discussed hereafter. 

We then speculate upon the expected peak flux density of H26$\alpha$ line 
to the observer ($F_{\rm H26\alpha,obs}$). 
Under the conditions described above, 
the intrinsic peak brightness temperature of 
H26$\alpha$ line is $\sim$ 0.5 $\times$ $T_e$ = 5000 K. 
Suppose that a single blackbody cloud 
(radius = 10 r$_\odot$) at this temperature is radiating, 
an expected peak flux density from the cloud is 
1.48 $\times$ 10$^{-14}$ Jy for the case of NGC 1068. 
As we have assumed that a single cloud has 
the thermal width of 21.5 km s$^{-1}$, 
an expected integrated intensity under the Gaussian line profile is then 
$S_{\rm H26\alpha,cl}$ = 3.38 $\times$ 10$^{-13}$ Jy km s$^{-1}$. 
This value will be multiplied by a number of BLR clouds 
contributing to the observed flux to estimate $F_{\rm H26\alpha,obs}$. 

To do so, we first estimate the effective source size of 
the line-emitting region in the BLR. 
Reverberation mapping observations towards type 1 AGNs revealed 
that the innermost radius of a dusty torus, $r_{\rm in}$, 
is proportional to the square root of the AGN luminosity (e.g., \citealt{2006ApJ...639...46S}). 
To estimate the $r_{\rm in}$ of NGC 1068, we first employed [\ion{O}{iv}] luminosity, 
which is considered to be an isotropic proxy of the AGN power (e.g., \citealt{2009ApJ...698..623D}), 
thus is more likely to be applicable to obscured AGNs. 
Using the compiled data in \citet{2009ApJ...698..623D}, 
we find that $r_{\rm in}$ of NGC 1068 would be $\sim$ 1.5 times 
larger than that of the type 1 Seyfert NGC 7469. 
Note that these AGNs have comparable (absorption corrected) 
X-ray luminosity as well (\citealt{2014ApJ...783..106L,2016MNRAS.456L..94M}), 
which supports the view that they have accretion discs with a similar condition. 
As the reverberation mapping by \citet{2014ApJ...788..159K} 
revealed that $r_{\rm in}$ of NGC 7469 is $\sim$ 100 light-days, 
we deduce that of NGC 1068 would be $\sim$ 150 light-days (0.13 pc). 
This is much larger than the expected blackbody radius of NGC 1068 for 10$^4$ K ($\la$ 0.005 pc), 
indicating net radiative excitation (absorption minus stimulated emission; \citetalias{2013ApJ...779...75S}) is insignificant for this BLR. 

For optically thick emission, an areal covering factor ($f_a$), 
rather than a volume-filling factor, is important to estimate the effective source size. 
So far, simple speculation on the equivalent width of Ly$\alpha$ 
yields $f_a$ $\sim$ 0.1 (\citealt{1997iagn.book.....P}) in BLRs. 
Therefore, the total H26$\alpha$ flux expected for NGC 1068 can be, 
\begin{equation}
   S_{\rm H26\alpha,cl} \times \frac{f_a \pi r^2_{\rm in}}{\pi (10 {\rm r}_\odot)^2} \sim 0.011 ~\rm{Jy~km~s^{-1}}, 
	\label{eq7}
\end{equation}
which is indeed below the 3$\sigma$ detection limit of our observations (4.02 Jy km s$^{-1}$ for the single beam; Table \ref{tbl1}). 
Although readers would scale this value with their own parameters, 
with this estimation, we consider that our non-detection 
of the broad H26$\alpha$ line is primarily due to our limited sensitivity. 
Furthermore, it is virtually impossible to achieve a high sensitivity 
enough to detect the above flux significantly (e.g., 10$\sigma$ detection) 
even with ALMA, as the expected peak flux density is $\sim$ a few $-$ 10 $\mu$Jy. 
On the other hand, we will achieve a 1$\sigma$ sensitivity\footnote{\url{https://almascience.nrao.edu/proposing/sensitivity-calculator}} 
of only $\sim$ 5 $\mu$Jy even with 100 days on-source integration at a velocity resolution of 100 km s$^{-1}$. 

When the electron density is high ($N_e$ $\ga$ 10$^{10}$ cm$^{-3}$), 
the situation becomes essentially different 
since both the H26$\alpha$ line and the underlying 
free$-$free continuum emissions are now optically thick. 
In this case, $R_{\rm lc}$ $\sim$ 0 is realized, 
whereas the continuum emission is maximally emitted as the blackbody. 
Hence, again one can never detect the line component 
from these extremely high-density clouds. 
For clouds with the low-end $N_e$ expected for BLRs ($\sim$ 10$^8$ cm$^{-3}$), 
where both the H26$\alpha$ line and the free$-$free emissions are optically thin, 
we find that $R_{\rm lc}$ $\sim$ 5. 
However, the very low optical depth ($\tau_{\rm H26\alpha}$ = 0.03) 
leads the emission to be totally transparent, 
which cannot be detected at all. 

\begin{figure}
	\includegraphics[width=\columnwidth]{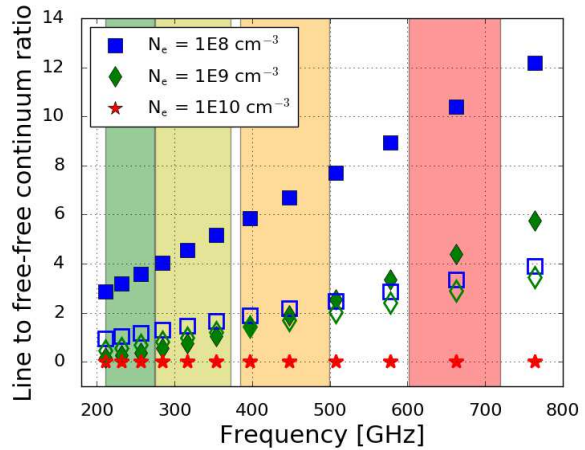}
    \caption{Expected line to free$-$free continuum ratios as a function of frequency. 
    Three different cases of $N_e$ are displayed for twelve submm-RLs (H31$\alpha$ to H20$\alpha$), 
    i.e., $N_e$ = 10$^8$ (blue-square), 10$^9$ (green-diamond), and 10$^{10}$ (red-star) cm$^{-3}$. 
    The filled and open symbols indicate that $T_e$ = 10$^4$ K and 2 $\times$ 10$^4$ K, respectively. 
    As we fixed $L$ = 20 r$_\odot$ in this calculation, $N_e$ = 10$^9$ cm$^{-3}$ 
    corresponds to $EM$ = 4.5 $\times$ 10$^{11}$ cm$^{-6}$ pc. 
    The shaded regions denote the frequency coverages of the ALMA bands, 
    from Band 6 (green) to Band 9 (red). 
    }
    \label{fig4}
\end{figure}

Under the LTE condition, we expand the above speculations 
to other submm-RLs that can be observed with ALMA (redshift = 0 is assumed here). 
The resultant $R_{\rm lc}$ of H31$\alpha$ to H20$\alpha$, 
as well as the frequency coverages of the ALMA Bands 6 to 9 are shown in Fig. \ref{fig4}. 
Lower frequency cases (e.g., H40$\alpha$) are not displayed 
since their $R_{\rm lc}$ approach to $\sim$ 0 for $N_e$ $\ga$ 10$^9$ cm$^{-3}$ due to large optical depths. 
We varied $N_e$ and $T_e$ whereas $L$ is fixed to 20 r$_\odot$. 
Thus, $N_e$ = 10$^9$ cm$^{-3}$ corresponds to EM = 4.5 $\times$ 10$^{11}$ cm$^{-6}$ pc. 
As already explained, one can see that $R_{\rm lc}$ $>$ 0 for $N_e$ = 10$^8$ cm$^{-3}$ and 10$^9$ cm$^{-3}$ cases 
but $R_{\rm lc}$ $\sim$ 0 for $N_e$ = 10$^{10}$ cm$^{-3}$ (or higher) case. 
We note that submm-RLs from $N_e$ = 10$^8$ cm$^{-3}$ clouds 
are totally optically thin and are hard to be detectable. 
Therefore, BLR clouds with $N_e$ $\sim$ 10$^9$ cm$^{-3}$ 
would be the dominant contributor to the H$n\alpha$ flux. 
Even so, however, we again find that the line fluxes of any H$n\alpha$ 
are intrinsically as extremely faint as calculated above, 
for virtually all imaginable sets of physical parameters expected in BLRs. 
Hence, we suppose that it is very impractical to detect submm-RLs of the BLR-origin even with ALMA. 

At centimetre wavelengths, on the other hand, one would expect to see BLR lines 
as absorption against bright synchrotron emission with $T_{\rm B}$ $\gg$ 10$^4$ K, 
if a synchrotron source (e.g., jet core) is embedded in the BLR, 
which is an analogous to observations of \ion{H}{i} absorption line at 21 cm (e.g., \citealt{2003ApJ...597..809M}). 
This detection requires a quite high resolution (e.g., sub-pc scale) 
because otherwise the surrounding jet component would contaminate the continuum level. 
While Very Long Baseline Interferometry systems 
seem to have this capability ($n$ $\ga$ 90 RLs shall be observed), 
their narrow bandwidths are eventually problematic in actual observations. 
Moreover, if the synchrotron emission in NGC 1068 has 
an electron-scattered nature as suggested by \citet{2011ApJ...736...37K}, 
the absorption feature would be outshone by that scattered light. 
Therefore, it would be quite difficult to detect the BLR components at centimetre bands. 

In the last of this article, we mention that a similar speculation in this work 
can be applied for the non-detection of H26$\alpha$ line of NLR-origin as well. 
Suppose that ionized gas with NLR-like parameters (e.g., $T_e$ = 10$^4$ K, $N_e$ = $N_p$ = 10$^4$ cm$^{-3}$) 
is filling a spherical volume within our observing beam ($\sim$ 24 pc in radius), 
Equations (11), (13), and (17) of \citetalias{2013ApJ...779...75S} yields an integrated H26$\alpha$ intensity 
to be $\sim$ 135 Jy km s$^{-1}$. 
This is $>$ 200 times larger than our 3$\sigma$ detection limit of 
$\sim$ 0.67 Jy km s$^{-1}$ (500 km s$^{-1}$ FWHM is assumed; \citealt{2011ApJ...739...69M}). 
Considering that submm-RLs of NLR-origin are likely to be optically thin (\citetalias{2013ApJ...779...75S}), 
this discrepancy can be largely attributed to a small volume-filling factor of ionizing clouds. 
From our observations, that factor is estimated to be $<$ 5 $\times$ 10$^{-3}$, 
which is consistent with the canonical value for NLRs, $<$ 10$^{-2}$ (\citealt{1997iagn.book.....P}).

%%%%%%%%%% SECTION 5: SUMMARY AND CONCLUSION %%%%%%%%%%
\section{Summary and Conclusions}\label{sec5}
In this work, we tried to detect a submm-RL of H26$\alpha$  
stemming from the heavily dust-obscured BLR of NGC 1068 
with high-resolution ALMA Cycle 2 observations. 
Our key results are as follows. 

\begin{itemize}
\item Admitting the uncertainty in the continuum level due to 
the limited bandwidth to fully cover the hypothesized line width, 
we found the non-detection of both broad and narrow H26$\alpha$ emission line 
towards the nucleus of NGC 1068 at the 3$\sigma$ sensitivity of 1.26 mJy at d$V$ = 50 km s$^{-1}$ 
(0.37 arcsec $\times$ 0.31 arcsec = 26 pc $\times$ 22 pc beam). 
Note that this non-detection is not conclusive 
when we adopt a lower continuum level, which is hard to determine accurately at this moment. 
\item We investigated the nature of BLR clouds 
that can potentially emit broad submm-RLs under the LTE condition. 
As a result, we suggest $N_e$ $\sim$ 10$^9$ cm$^{-3}$ components 
mostly contributes to line emission. 
From clouds with other densities, we will not see broad submm-RLs due to too large or too small optical depths. 
$N_e$ $\sim$ 10$^9$ cm$^{-3}$ is typical for BLRs. 
However, an expected line flux even for this $N_e$ = 10$^9$ cm$^{-3}$ case is extremely faint based on our calculation. 
Hence, our non-detection of broad H26$\alpha$ is due to the limited sensitivity. 
We note that one can virtually never achieve high enough sensitivity to detect such faint emission even with ALMA. 
\item We predicted a line-to-continuum ratio of various H$n\alpha$ at submillimetre 
to investigate whether the line components can appear. 
As a result, although some combinations of parameters can yield the broad line component in principle, 
again we cannot detect them due to the limited sensitivity of currently available submillimetre instruments. 
\end{itemize}

Considering the discussion in Section \ref{sec4}, 
as well as the limited column penetrating capability at NIR wavelength, 
one possible way to directly detect obscured BLR lines 
is to employ sensitive mid-IR observations. 
Such a chance will be provided with the Mid-infrared Instrument (MIRI)\footnote{\url{http://jwst.nasa.gov/miri.html}} 
on board the {\it{James Webb Space Telescope (JWST)}}\footnote{\url{http://www.jwst.nasa.gov}}. 
That instrument can cover $n$ = 5--8 \ion{H}{i} RLs at redshift zero, 
with the 10 $\sigma$ detection limit of $\sim$ 1 $\times$ 10$^{-20}$ W m$^{-2}$ 
for a point source (at $\sim$ 10 $\mu$m, with 10 000 s). 
However, careful treatment on the departure from the LTE as well as the influence 
from both thermal (from an accretion disc and a dusty torus) 
and non-thermal continuum radiation is necessary to predict the detectability.

\section*{Acknowledgements}
We thank the anonymous referee for his/her kind and encouraging suggestions to improve this article. 
This paper makes use of the following ALMA data: ADS/JAO.ALMA\#2013.1.00188.S. 
ALMA is a partnership of ESO (representing its member states), NSF (USA), and NINS (Japan), 
together with NRC (Canada) and NSC and ASIAA (Taiwan), 
in cooperation with the Republic of Chile. 
The Joint ALMA Observatory is operated by ESO, AUI/NRAO, and NAOJ. 
The National Radio Astronomy Observatory is a facility of the National Science Foundation 
operated under cooperative agreement by Associated Universities, Inc. 
KN acknowledges support from JSPS KAKENHI Grant Number 15K05035. 
TI is thankful for the fellowship received from the Japan Society for the Promotion of Science (JSPS).

%%%%%%%%%%%%%%%%%%%%%%%%%%%%%%%%%%%%%%%%%%%%%%%%%%

%%%%%%%%%%%%%%%%%%%% REFERENCES %%%%%%%%%%%%%%%%%%
% The best way to enter references is to use BibTeX:
\bibliographystyle{mnras}
\bibliography{reference} % if your bibtex file is called example.bib

%%%%%%%%%%%%%%%%%%%%%%%%%%%%%%%%%%%%%%%%%%%%%%%%%%

%%%%%%%%%%%%%%%%% APPENDICES %%%%%%%%%%%%%%%%%%%%%

\appendix
%\section{Some extra material}
% SOMETHING TO WRITE HERE ???

%%%%%%%%%%%%%%%%%%%%%%%%%%%%%%%%%%%%%%%%%%%%%%%%%%

% Don't change these lines
\bsp	% typesetting comment
\label{lastpage}
\end{document}